\documentclass{moriond}

\usepackage{amsmath}
\def\Journal#1#2#3#4{{#1} {\bf #2}, #3 (#4)}

\def\PLB{{\em Phys. Lett.}  B}
\def\PRL{\em Phys. Rev. Lett.}
\def\PRD{{\em Phys. Rev.} D}

\def\EPJC{{\em Eur. Phys. J.} C}
\def\JHEP{{\em JHEP}}
\def\JINST{{\em JINST}}
\def\be{\begin{equation}}
\def\ee{\end{equation}}
\def\bea{\begin{eqnarray}}
\def\eea{\end{eqnarray}}

\newcommand{\Photo}{\includegraphics[height=35mm]{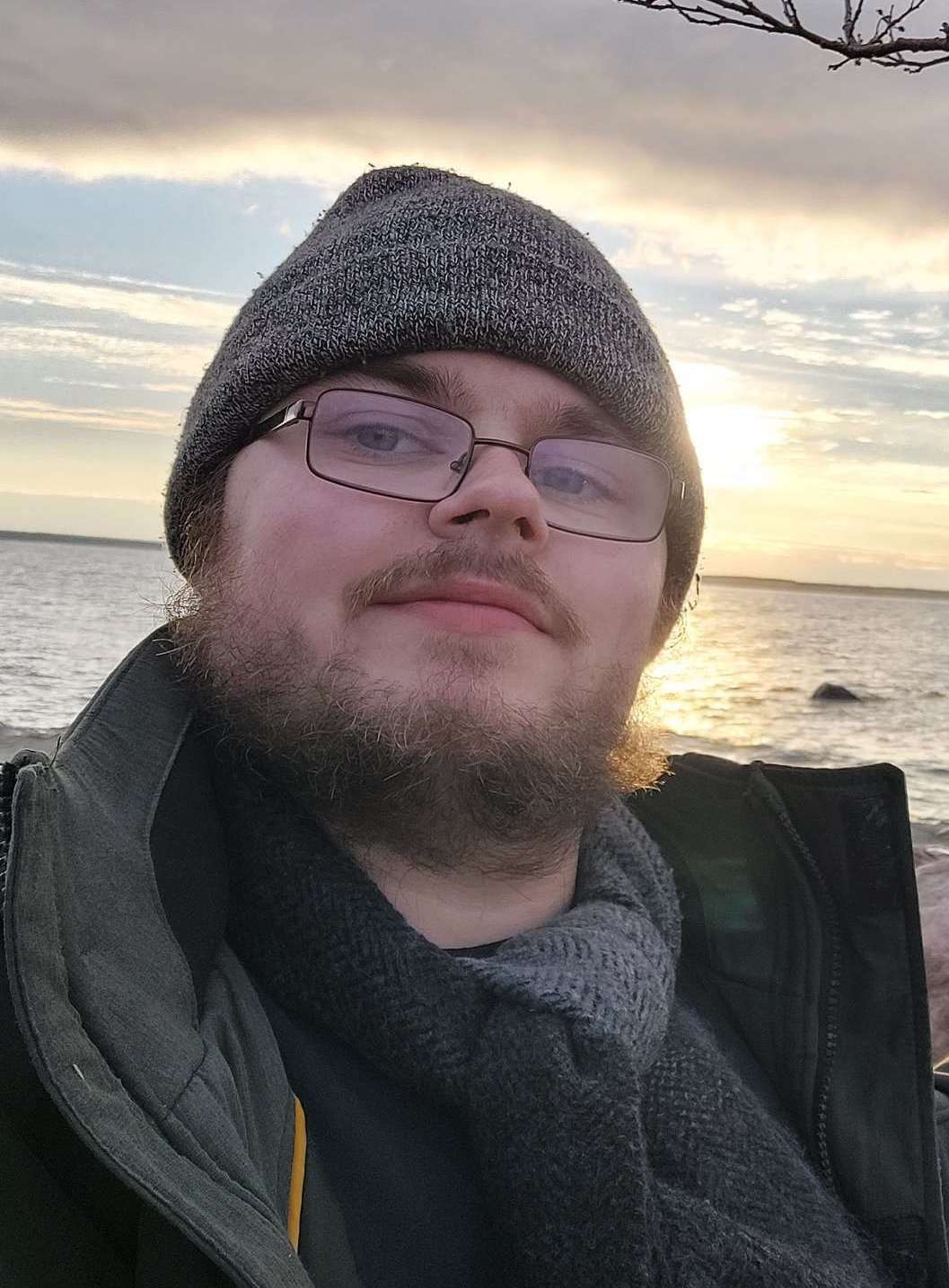}}

\begin{document}
\vspace*{4cm}
\title{Higgs cross-section (including di-Higgs) with CMS and ATLAS
}

\author{T. Lange on behalf of the ATLAS and CMS Collaborations }

\address{National Institute of Chemical Physics and Biophysics (KBFI/NICPB), KEAL, Rävala pst 10,\\
10145 Tallinn, Estonia}

\maketitle\abstracts{
Since the discovery of the Higgs boson in 2012 by the ATLAS and CMS Collaborations a lot of progress has been made in verifying the nature of this new bosonic particle. Still, questions remain as to whether this new particle is the standard model (SM) Higgs boson, and whether it gives us hints of where physics beyond the SM (BSM) might be hidden. This overview tries to tackle these questions by looking at three types of analyses 1) Total cross section measurements of the various Higgs boson production and decay modes, allowing us access to the various Higgs boson couplings to SM bosons and fermions, 2) Differential Higgs boson cross section measurements such as in the STXS framework allowing for a model independent search for BSM and, finally, 3) The searches for di-Higgs production which give access to the trilinear Higgs self coupling $\lambda$ and the Higgs potential itself.}

\section{Introduction}
The Higgs boson discovery by the ATLAS~\cite{ATLAS:2008xda} and CMS~\cite{CMS:2008xjf} Collaborations~\cite{ATLAS:2012yve,CMS:2012qbp,CMS:2013btf} was one of the last centuries biggest achievements in high energy physics. Great progress has been made in studying this new boson, but the overall question remains, is this the Standard Model (SM) predicted Higgs boson? Or does this new particle maybe give us hints at physics beyond the SM (BSM)? To answer this question, the ATLAS and CMS Collaborations study every facet of the Higgs boson with ever increasing scrutiny. In the context of the overview given in this talk, three aspects are investigated:
\textbf{1)} Measurements of total cross sections for the various production and decay modes of the Higgs boson. These measurements give us direct access to the Higgs bosons couplings to other SM particles.
\textbf{2)} Measurements of differential Higgs boson cross sections such as in the STXS 1.2 framework~\cite{LHCHiggsCrossSectionWorkingGroup:2016ypw}, allowing for the constraint of BSM physics in a model-independent way.
\textbf{3)} Measurements of di-Higgs boson production, giving us access to so far inaccessible Higgs couplings, such as the trilinear Higgs boson  self-coupling ($\lambda$) with a direct link to the Higgs boson potential itself, or the quartic coupling between two Higgs bosons and two vector bosons ($C_{2V}$).
This overview is not an exhaustive one, after a very successful LHC Run 2 data taking period, more Higgs boson measurements have been performed than can be covered here.
\section{Single-Higgs boson analyses}
\begin{figure}\centering
\begin{minipage}{0.36\linewidth}
\centerline{\includegraphics[width=0.85\linewidth]{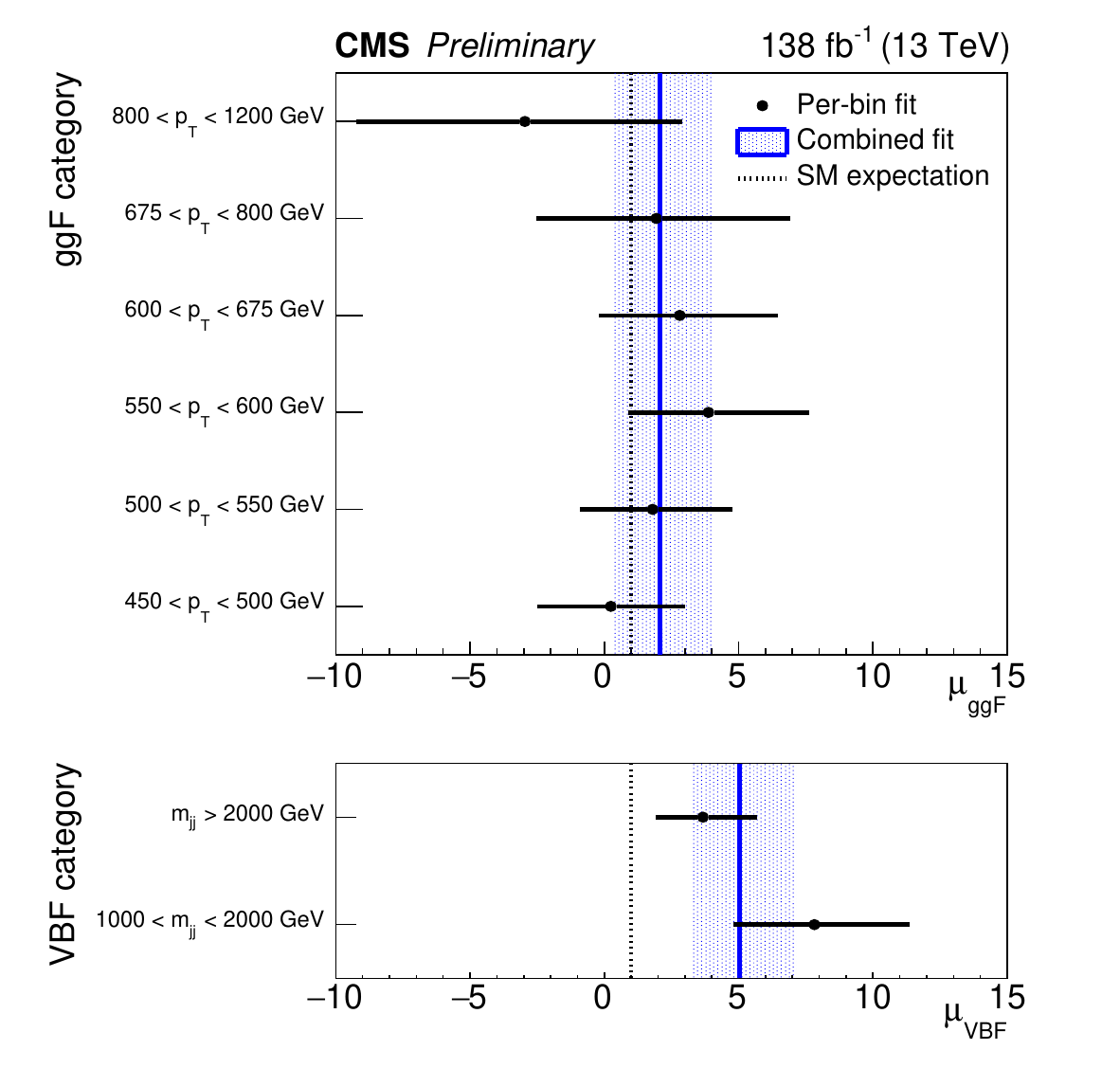}}
\end{minipage}
\hspace{0.1\textwidth}
\begin{minipage}{0.44\linewidth}
\centerline{\includegraphics[width=\linewidth]{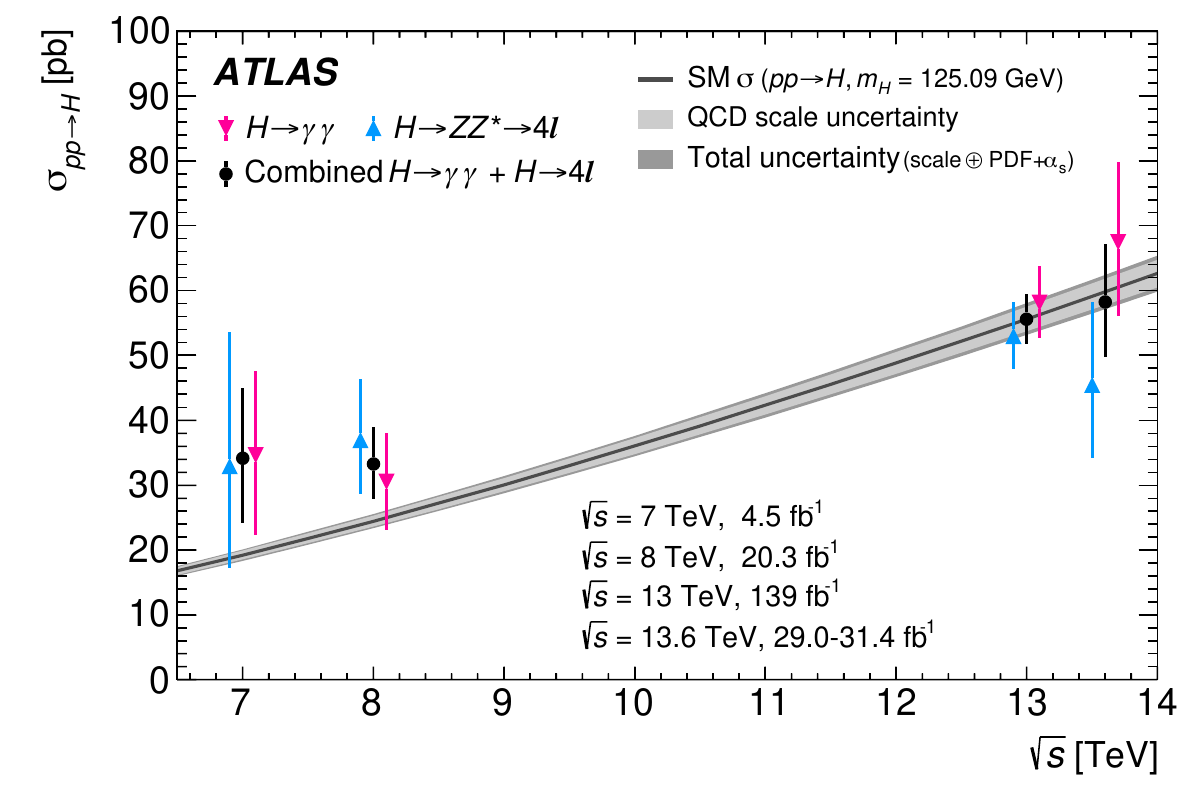}}
\end{minipage}
\hfill
\caption[]{Illustrative figures for the boosted $H\rightarrow b\bar{b}$ search~\cite{CMS-PAS-HIG-21-020} (left), and the  ATLAS measurement~\cite{ATLAS:2023tnc} at $\sqrt{s}$=13.6 TeV for $H\rightarrow \gamma\gamma$ and $H\rightarrow ZZ^*$ (right).}
\label{fig:oneTothree}
\end{figure}

\begin{figure}\centering
\begin{minipage}{0.375\linewidth}
\centerline{\includegraphics[width=\linewidth]{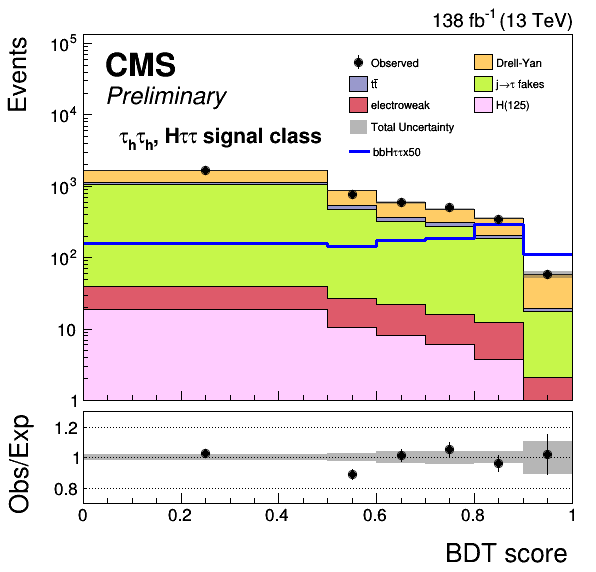}}
\end{minipage}
\hspace{0.1\textwidth}
\begin{minipage}{0.33\linewidth}
\centerline{\includegraphics[width=\linewidth]{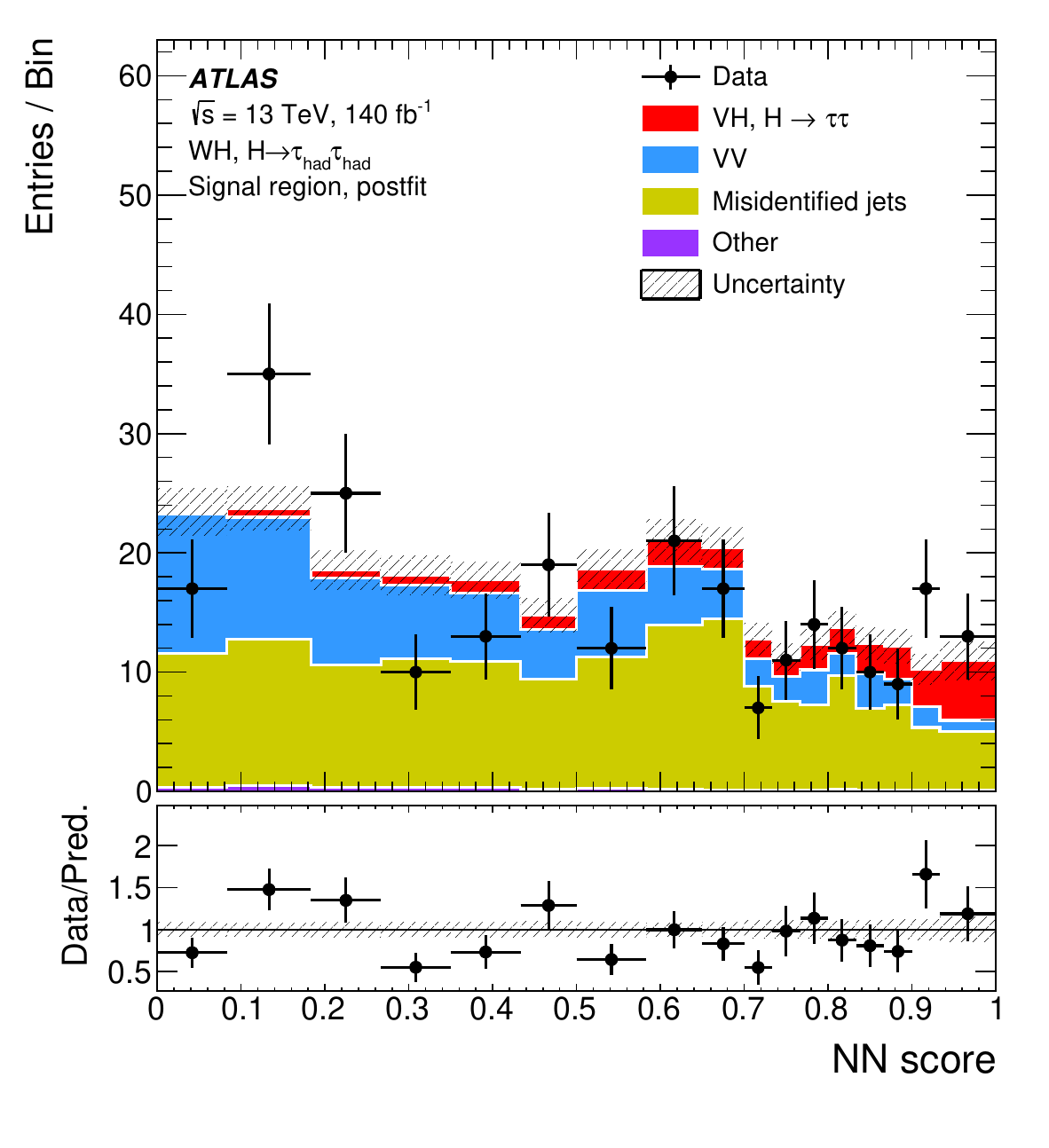}}
\end{minipage}
\hfill
\caption[]{Illustrative figures for the CMS $bbH\rightarrow WW^*/\tau\tau$ analysis~\cite{CMS-PAS-HIG-23-003} (left) and the ATLAS $H\rightarrow \tau\tau$ in the $VH$ production mode~\cite{ATLAS:2023qpu} (right).}
\label{fig:fourTofive}
\end{figure}

$H\rightarrow b\bar{b}$: Starting with the Higgs boson decay with the highest branching fraction, several new analyses have been brought forward by the ATLAS and CMS Collaborations~\cite{CMS-PAS-HIG-21-020,ATLAS:2023jdk,CMS:2023vzh,CMS-PAS-HIG-19-011}. Of particular interest is the CMS search for boosted $H\rightarrow b\bar{b}$ decays~\cite{CMS-PAS-HIG-21-020} providing the first search of $H\rightarrow b\bar{b}$ in the VBF production mode. This analysis makes use of an improved version of the \textbf{DeepDoubleX} tagger~\cite{CMS-DP-2022-041} used to identify the $H\rightarrow b\bar{b}$ decay (DDB). Important backgrounds of this analysis are given by QCD multijet production, estimated from a DDB sideband, and $t\bar{t}$ production, normalized with the help of a muon control region included in the final likelihood fit on the invariant $H\rightarrow b\bar{b}$ mass. The measured cross sections are shown in Fig.~\ref{fig:oneTothree} (left).\\
$H\rightarrow Z^*\gamma$: In contrast to $H\rightarrow b\bar{b}$ with one of the highest Higgs boson branching fractions, $H\rightarrow Z^*\gamma$ has one of the smallest. Despite this the ATLAS and CMS Collaborations where able to provide the first evidence for this rare SM process~\cite{ATLAS:2023yqk}. The signal is extracted by searching for a peaking signal on top of the smoothly falling background spectrum in $m_{\ell^+\ell^-\gamma}$. A best fit signal strength $\hat{\mu}$ of $2.0^{+1.0}_{-0.9}$/$2.4^{+1.0}_{-0.9}$ is extracted for ATLAS/CMS respectively, with the combination yielding $\hat{\mu}=2.2\pm 0.6\text{(stat.)}^{+0.3}_{-0.2}\text{(syst.)}$.\\
$H\rightarrow \gamma\gamma$ and $H\rightarrow ZZ^*$: The first LHC Run3 Higgs result is shown in~\cite{ATLAS:2023tnc}. While this result does not introduce new analysis strategies, this result not only shows that the new ATLAS muon detector works as anticipated, but it continues to investigate the Higgs boson from a new angle: the energy frontier. As shown in the right part of Fig.~\ref{fig:oneTothree}, there are now Higgs boson cross section measurements at a total of four LHC center of mass energies, showing agreement with the SM over the whole energy range of the LHC.\\
$bbH\rightarrow WW^*/\tau\tau$:
CMS is presenting its first Run 2 search for $bbH$ production~\cite{CMS-PAS-HIG-23-003}. This analysis targets the non Higgs Strahlungs production modes of $bbH$ in four final states with different multiplicities in hadronic taus ($\tau_{h}$) and electrons/muons. Important backgrounds for this search are $t\bar{t}$ and $V+\text{jets}$ production, QCD multijet backgrounds and backgrounds from misidentified $\tau_h$. Backgrounds with misidentified $\tau_h$ candidates are estimated from data. The final 95\% CL upper limit of 3.7 $\times$ SM cross section (6.1 expected) is extracted using a multi class BDT with one example distribution shown in the left part of Fig.~\ref{fig:fourTofive}.\\
$H\rightarrow \tau\tau$ and ATLAS combination:
ATLAS presents a new search for $H\rightarrow \tau\tau$ in the $VH$ production mode focusing on leptonically decaying $W$ and $Z$ bosons~\cite{ATLAS:2023qpu}. The results are extracted in a likelihood fit to outputs of 6 neural networks (NN) in 4 categories based on the multiplicities of $\tau_h$, electrons and muons in the final state. Similar to other analyses with $\tau_h$, the important background from misidentified $\tau_h$ is estimated from data using the Fake Factor method with other backgrounds taken from simulation. This analysis shows a measurement for $H\rightarrow \tau\tau$ in the VH production mode at 4.2 $\sigma$ (3.6 $\sigma$ expected). The right part of Fig.~\ref{fig:fourTofive} shows an example NN output. More results from ATLAS can be found in a new combined measurement~\cite{ATLAS:2024lyh} which not only features a good overview of all ATLAS differential Higgs boson measurements but also many BSM reinterpretations.

\section{Double Higgs Analysis}
\begin{figure}\centering
\begin{minipage}{0.42\linewidth}
\centerline{\includegraphics[width=\linewidth]{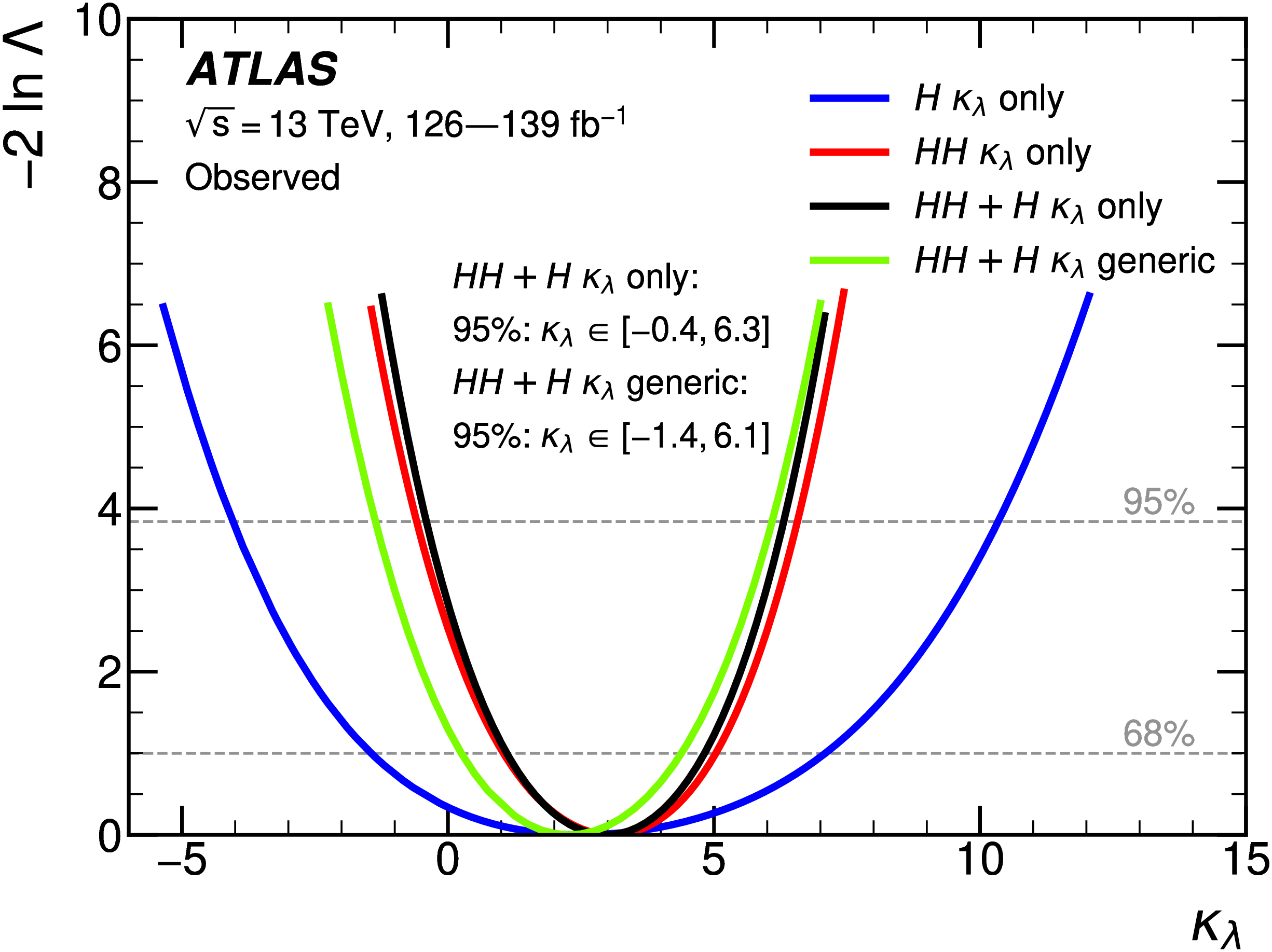}}
\end{minipage}
\hspace{0.1\textwidth}
\begin{minipage}{0.39\linewidth}
\centerline{\includegraphics[width=\linewidth]{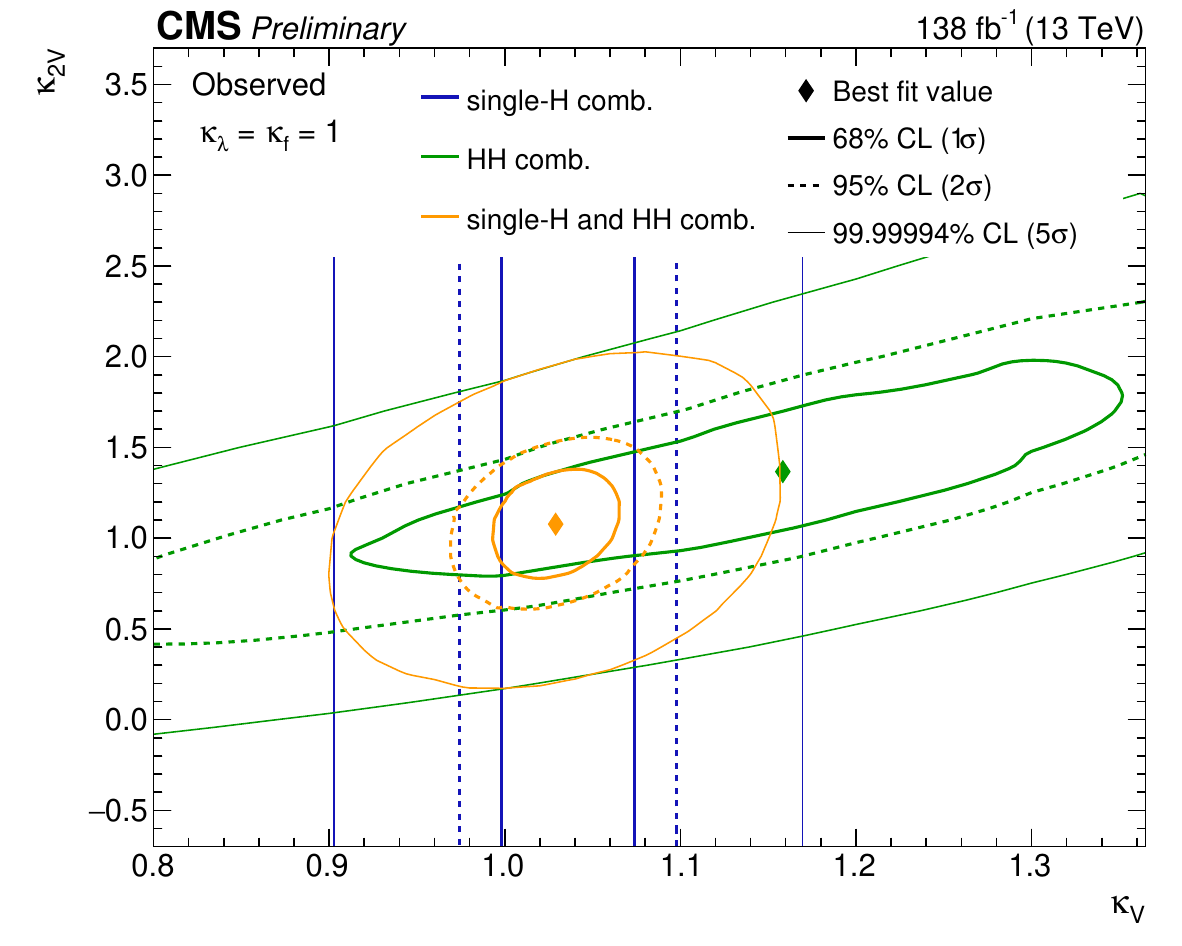}}
\end{minipage}
\hfill
\caption[]{Illustrative figures giving an overview about published di-Higgs searches from ATLAS and CMS in Run 2 (left), results from an ATLAS~\cite{ATLAS:2022jtk} (middle) and CMS~\cite{CMS-PAS-HIG-23-006,ATLAS:2022jtk} (right) H+HH combination. }
\label{fig:last}
\end{figure}

The production of Higgs pairs (di-Higgs) represents one of the rarest SM processes~\cite{LHCHiggsCrossSectionWorkingGroup:2016ypw} with with a cross section of just 31 fb due to the destructive interference of two processes in its main production mode $ggHH$. Many different decay channels are possible in the analysis of HH production; ATLAS and CMS explore several of them to maximise sensitivity to this rare SM process. Individual analyses are now reaching sensitivities up to O(3-5) the SM rate, with combinations reaching sensitivity to O(2-3) the SM rate~\cite{CMS:2022dwd,ATLAS-CONF-2021-052}. These analyses give access to unique Higgs couplings, such as the Higgs boson self-coupling $\lambda$ with a direct link to the Higgs boson potential itself, or the quartic coupling between two Higgs bosons and two vector bosons $C_{2V}$.\\
$HH\rightarrow \tau\tau\gamma\gamma$: CMS presents the first search~\cite{CMS-PAS-HIG-22-012} for $HH\rightarrow \tau\tau\gamma\gamma$. This new channel achieves a 95\% CL limit on di-Higgs production of 33 (26 expected) $\times$ SM rate. This result is obtained by searching for a signal peak on top of the background in the diphoton invariant mass spectrum, in eight different categories based on the multiplicity of $\tau_h$, electrons and muons.\\
$HH\rightarrow 4b$ (boosted): Another new result, is the first search by the ATLAS Collaboration for boosted $HH\rightarrow 4b$ decays~\cite{ATLAS-CONF-2024-003}. Similar to the corresponding result by CMS, this analysis is especially sensitive to the $C_{2V}$ coupling, excluding $C_{2V}=0$ at 3.8 $\sigma$ when fixing all other Higgs boson couplings to their SM values. This result is extracted by a likelihoodfit to the distribution of a BDT in the signal region defined by the use of a double-b tagger~\cite{ATL-PHYS-PUB-2020-019} identifying the $H\rightarrow 4b$ decay and a mass window in the invariant masses of the two b-quark pairs of the final state. Other new results by ATLAS also include a search~\cite{ATLAS:2023elc} for $HH\rightarrow bb+\ell\ell+p_T^\text{miss}$ and a search~\cite{ATLAS-CONF-2024-005} for $HH\rightarrow\text{leptons/photons}$. \\
$H+HH$ combinations: ATLAS and CMS also published their first results for combinations of single H and di-Higgs searches~\cite{CMS-PAS-HIG-23-006,ATLAS:2022jtk}. These results are especially useful to investigate the di-Higgs related couplings $\lambda$ and $C_{2V}$ without making assumptions on other Higgs boson couplings, as shown in the $\lambda$ scan in the left part of Fig.~\ref{fig:last} and the two dimensional scan of $C_{2V}$ and $C_V$ shown in the right of Fig.~\ref{fig:last}, excluding $C_{2V}=0$ for any value of $C_V$.
\section{Summary and Conclusion}
Higgs results for the LHC Run 3 are ramping up now after a very successful LHC Run 2 with many results investigating every facet of the Higgs boson. These results include cross section measurements, searches for rare Higgs boson production and decay modes, differential measurements, and searches for di-Higgs production, which are progressively approaching SM sensitivity.
\section*{Acknowledgments}
I would like to thank the Moriond organizers and the ATLAS and CMS Collaborations for the opportunity to present this work at the 58th Rencontres de Moriond. I would also like to thank the Estonian Research council for their support with the CoE grant TK202 “Foundations of the Universe” and the CERN Science Consortium of Estonia grant RVTT3. I would also like to thank the COST action CA22130 COMETA to which part of the presented work is related.

\end{document}